\documentclass[a4paper,11pt]{article}
\usepackage{pos}
\usepackage{slashed}
\title{Connecting Seesaw Effective Field Theory to Full Theory via Flavor Invariants}
\ShortTitle{Connecting SEFT to Full Theory via Flavor Invariants}

\author*[a,b]{Bingrong Yu}
\author[a,b]{Shun Zhou}

\affiliation[a]{Institute of High Energy Physics, Chinese Academy of Sciences,\\
Beijing 100049, China}

\affiliation[b]{School of Physical Sciences, University of Chinese Academy of Sciences,\\
Beijing 100049, China}

\emailAdd{yubr@ihep.ac.cn}
\emailAdd{zhoush@ihep.ac.cn}

\abstract{The canonical seesaw models are one of the simplest and most natural scenarios that can account simultaneously for neutrino masses and  matter-antimatter asymmetry in our universe. Below the seesaw scale, one can integrate out the heavy degrees of freedom to construct the seesaw effective field theory (SEFT). In this talk, we investigate the connection between the full seesaw model and the low-energy SEFT from a brand-new perspective: the invariant theory. Using the powerful tool of Hilbert series, we demonstrate the intimate relation between the flavor space of the SEFT and that of its ultraviolet theory. Through the matching of flavor invariants, it is revealed that the precise measurements of dimension-five and dimension-six operators in the SEFT at low energies are powerful enough to probe the full seesaw model, including CP violation necessary for successful leptogenesis.
}

\FullConference{%
  41st International Conference on High Energy physics - ICHEP2022\\
  6-13 July, 2022\\
  Bologna, Italy
}

\begin{document}
\maketitle

\section{Motivation}
\vspace{-0.2cm}
The type-I seesaw model~\cite{typeI}, which extends the Standard Model (SM) by adding right-handed (RH) neutrinos $N_{\rm R}^{}$, can naturally explain the tiny neutrino masses and generate the cosmological matter-antimatter asymmetry through leptogenesis~\cite{Fukugita:1986hr}.
In the type-I seesaw model, the RH neutrinos are usually much heavier than the electroweak scale and thus difficult to be directly observed in the collider experiments. In this case, it will be practically useful to integrate out the heavy degrees of freedom to obtain the low-energy effective theory of the seesaw model, the so-called seesaw effective field theory (SEFT), which governs all the low-energy phenomena of the full seesaw model. Then one may immediately ask: How does the ultraviolet (UV) theory affect the observables at low energies? And conversely, how much can we know about the full seesaw model from the low-energy experiments?

From the viewpoint of effective field theory, the impact of the UV theory on low-energy observables is completely encoded in the Wilson coefficients of high-dimensional effective operators. The connection between the full seesaw model and the SEFT can be established by the parameter matching at the seesaw scale $\Lambda$.
The matching of the type-I seesaw model onto the SEFT up to ${\cal O}(1/\Lambda^2)$ at the tree level induces the dimension-five Weinberg operator ${\cal O}^{\alpha \beta}_5 = \overline{\ell^{}_{\alpha \rm L}} \widetilde{H} \widetilde{H}^{\rm T} \ell^{\rm C}_{\beta \rm L}$~\cite{Weinberg:1979sa} and one dimension-six operator ${\cal O}^{\alpha \beta}_6 = \left(\overline{\ell^{}_{\alpha \rm L}} \widetilde{H}\right) {\rm i}\slashed{\partial}\left( \widetilde{H}^\dagger \ell^{}_{\beta \rm L}\right)$ ~\cite{dim6}. Although there have been many efforts in studying the relations between the observables in the low-energy SEFT and those in the full seesaw model ~\cite{dim6,bridge}, we will tackle this problem from a completely new point of view: the invariant theory~\cite{textbooks}. See Ref.~\cite{CPinvariant}
for early applications of (flavor) invariants to describing CP violation in and beyond the SM, and Refs.~\cite{Jenkins,Yu2021} for systematic studies of flavor structure in the quark and leptonic sector in the framework of the invariant theory.

\vspace{-0.2cm}
\section{Formalism}
\vspace{-0.2cm}
The relevant part of the Lagrangian in the  type-I seesaw is given by
\begin{eqnarray}
	\label{eq:full lagrangian}
	{\cal L}_{\rm seesaw}^{} = \overline{N^{}_{\rm R}}{\rm i}\slashed{\partial} N^{}_{\rm R} - \left[\overline{\ell_{\rm L}^{}}Y_\nu^{}\widetilde{H}N_{\rm R}^{} + \frac{1}{2}\overline{N_{\rm R}^{\rm C}}M_{\rm R}^{}N_{\rm R}^{}+{\rm h.c.} \right] \;, \quad
\end{eqnarray}
where ${\ell }_{\rm L}^{}$ and $\widetilde{H}\equiv {\rm i}\sigma_2^{}H_{}^{*}$ are the left-handed lepton doublet and the Higgs doublet, respectively. In addition, $Y_\nu^{}$ denotes the Dirac neutrino Yukawa coupling matrix and $M_{\rm R}^{}$ is the Majorana mass matrix of RH neutrinos. 
If the seesaw scale $\Lambda = {\cal O}(M^{}_{\rm R})$ is much higher than the electroweak scale, 
then the low-energy phenomena are determined by the SEFT Lagrangian
\begin{eqnarray}\label{eq:Left}
	\mathcal{L}^{}_{\rm SEFT} = \mathcal{L}^{}_{\rm SM} - \left[ \frac{C^{}_5}{2\Lambda} {\cal O}^{}_5 + {\rm h.c.} \right] + \frac{C^{}_6}{\Lambda^2} {\cal O}^{}_6 \; ,
\end{eqnarray}
up to ${\cal O}\left(1/\Lambda_{}^2\right)$. At the tree-level matching, the  Wilson coefficients are given by
\begin{eqnarray}
	\label{eq:wilson coe}
	C_5^{}=-Y_\nu^{}Y_{\rm R}^{-1}Y_\nu^{\rm T}\;, \quad
	C_6^{}=Y_\nu^{} \left(Y_{\rm R}^{\dagger}Y_{\rm R}^{}\right)_{}^{-1}Y_\nu^\dagger\;,\quad
	{\rm with}\;Y_{\rm R}^{}\equiv M_{\rm R}^{}/\Lambda\;. 
\end{eqnarray}
Now consider the most general transformation in the flavor space of the leptonic sector
\begin{eqnarray}
	\label{eq:field trans}
\ell_{\rm L}^{}\to U_{\rm L}^{}\ell_L^{}\;,\quad
l_{\rm R}^{}\to V_{\rm R}l_{\rm R}^{}\;,\quad
N_{\rm R}^{}\to U_{\rm R}^{}N_{\rm R}^{}\;, 
\end{eqnarray}
where $l_{\rm R}^{}$ denotes the RH charged-lepton fields. Here $U_{\rm L}^{}, V_{\rm R} \in {\rm U}(m)$ and $U_{\rm R}^{} \in {\rm U}(n)$ are three arbitrary unitary matrices (for $m$ lepton doublets and $n$ RH neutrinos). 
\newpage

\noindent Eq.~(\ref{eq:full lagrangian}) is invariant under the above flavor transformation only if the Yukawa matrices transform as  
\begin{eqnarray}
	\label{eq:Yukawa trans}
	Y_l^{} \to U_{\rm L}^{}Y_l^{}V_{\rm R}^\dagger\;,\quad
	Y_\nu^{} \to  U_{\rm L}^{}Y_\nu^{}U_{\rm R}^\dagger\;,\quad
	Y_{\rm R}^{} \to  U_{\rm R}^* Y_{\rm R}^{}U_{\rm R}^\dagger\;, 
\end{eqnarray}
where $Y_l^{}$ represents the Yukawa mass matrix of charged leptons. The transformation of the Wilson coefficients in the SEFT is induced by Eq.~(\ref{eq:Yukawa trans}) at the matching scale
\begin{eqnarray}
\label{eq:wilson coe trans}
C_5^{}\to U_{\rm L}^{}C_5^{}U_{\rm L}^{\rm T}\;,\quad
C_6^{}\to U_{\rm L}^{}C_6^{}U_{\rm L}^\dagger\;.
\end{eqnarray}
From Eqs.~(\ref{eq:Yukawa trans})-(\ref{eq:wilson coe trans}) we know that in the SEFT, the building blocks of the flavor invariants are $\{X_l^{} \equiv Y^{}_l Y^\dagger_l, C_5^{},C_6^{}\}$ with the symmetry group ${\rm U}(m)$, whereas $\{Y_l^{},Y_\nu^{},Y_{\rm R}^{}\}$ serve as the building blocks in the full seesaw model with the symmetry group ${\rm U}(m)\otimes {\rm U}(n)$.

Since the flavor invariants are closed under the addition and multiplication, they form a \emph{ring}. According to the invariant theory~\cite{textbooks}, all information about the algebraic structure of the invariant ring is encoded in the generating function of the ring, that is, the \emph{Hilbert series} (HS)
\begin{eqnarray}
{\cal H}(q)=\sum_{k=0}^{\infty}c_k^{}q_{}^k\;,
\end{eqnarray}
where $c_k^{}$ represent the number of linearly-independent invariants at degree $k$, $q$ is an arbitrary complex number to label the degrees of building blocks and satisfying $\left|q\right|<1$. 
See, e.g., Refs.~\cite{Jenkins,Yu2021} for more details about the HS and its applications in flavor physics. A systematic method to compute the HS is to use the Molien-Weyl (MW) formula~\cite{Molien-Weyl}.
As long as the symmetry group and the representations of the building blocks are given, the MW formula could reduce the computation of the HS to calculating contour integrals, which can be accomplished via the residue theorem~\cite{Jenkins,Yu2021}. 

\renewcommand\arraystretch{1.0}
\begin{table}[t!]
	\centering
	\begin{tabular}{l|c|c|c|c}
		\hline\hline
		{\bf models} & {\bf moduli} & {\bf phases} & {\bf physical parameters} & {\bf primary invariants}\\
		\hline
		2-generation SEFT ($m=2$) & 8 & 2 & 10 & 10\\
		\hline
		2-generation seesaw ($m=n=2$) & 8 & 2 & 10 & 10\\
		\hline
		3-generation SEFT ($m=3$) & 15 & 6 & 21 & 21\\
		\hline
		3-generation seesaw ($m=n=3$) & 15 & 6 & 21 & 21\\
		\hline\hline
	\end{tabular}
	\caption{\label{table:comparison} Comparison of the number of independent physical parameters and primary invariants  between the SEFT and the full seesaw model. Note that the moduli denote the parameters in the model other than phases.}
\end{table}
\renewcommand\arraystretch{1.0}

\vspace{-0.2cm}
\section{Results and Discussions}
\vspace{-0.2cm}
Now we apply the above general formalism of the invariant theory to our case. The symmetry groups in the flavor space of the SEFT and the full seesaw model are ${\rm U}(m)$ and ${\rm U}(m)\otimes{\rm U}(n)$, respectively. The representations of the building blocks under the symmetry groups are given by Eqs.~(\ref{eq:Yukawa trans})-(\ref{eq:wilson coe trans}). Using the MW formula, it is straightforward to compute the HS in the SEFT and the full theory~\cite{Yu2022}. From the HS one can explicitly construct all the basic and primary invariants and draw a number of very interesting conclusions (see Ref.~\cite{Yu2022} for more details):
\vspace{-0.2cm}
\begin{itemize}
\item There are exactly equal number of independent physical parameters and primary invariants in the SEFT and in its UV theory (cf. Table~\ref{table:comparison}).
This implies the inclusion of ${\cal O}_5^{}$ and ${\cal O}_6^{}$ in the effective theory is already \emph{adequate} to contain all physical information about
the full theory.
\item There are also equal number of generators (basic invariants) in the ring of the SEFT and that of the full seesaw model. There exists a one-to-one correspondence between two sets of CP-odd basic invariants. Therefore, CP conservation in the low-energy SEFT is equivalent to the absence of CP violation in the full theory.
\item Through a proper matching procedure of flavor invariants in the SEFT and in the full theory, one can directly relate CP violation for cosmological matter-antimatter asymmetry to that in low-energy oscillation experiments in a basis- and parametrization-independent way.
\end{itemize}
 
Through the language of invariant theory, it is quite clear that how the precise measurements at low energies can be used to probe the full seesaw model at high energies.

\vspace{-0.2cm}
\section*{Acknowledgements}
\vspace{-0.2cm}
This work was supported by the National Natural Science Foundation of China under grant No. 11835013.

\end{document}